\documentclass[a4paper]{jpconf}
\usepackage{graphicx}
\usepackage{amsmath}

\usepackage{color}
\usepackage{colordvi}

\definecolor{mygreen}{rgb}{0,0.5,0}

\def\eps{\epsilon}
\def\NGluon{{\tt NGluon}~}

\begin{document}
\pagestyle{plain}

\title{%
 \vskip-4\baselineskip%
 {\normalsize HU-EP-11/56\\%
   \normalsize SFB/CPP-11-74}%
 \vskip-3\baselineskip%
 \vskip4\baselineskip
Numerical evaluation of one-loop QCD amplitudes
}

\author{Simon Badger$^{1}$, Benedikt Biedermann$^{2}$ and Peter Uwer$^{2}$}
\address{%
$^{1}$ The Niels Bohr International Academy and Discovery Center, %
The Niels Bohr Institute, %
Blegdamsvej 17, DK-2100 Copenhagen, Denmark%
}
\address{%
$^{2}$ Humboldt-Universit\"at zu Berlin, Institut f\"ur Physik, %
Newtonstra{\ss}e 15, D-12489 Berlin, Germany%
}
\ead{benedikt.biedermann@physik.hu-berlin.de}

\begin{abstract}
We present the publicly available program \NGluon allowing the numerical
evaluation of primitive amplitudes at one-loop order in massless QCD. The
program allows the computation of one-loop amplitudes for an arbitrary number of
gluons. The focus of the present article is the extension
to one-loop amplitudes including an arbitrary number of massless quark
pairs. We discuss in detail the algorithmic differences to the pure
gluonic case and present cross checks to validate our
implementation. The numerical accuracy is investigated in detail.
\end{abstract}

\section{Introduction}

The automation of next-to-leading (NLO) corrections to multi-particle
processes in the Standard Model is an important step in making precision
predictions for signal and
  background reactions studied at the Large Hadron Collider (LHC)
at CERN. Recent years have seen considerable progress in simplifying this
complex task into a definite algorithm for arbitrary processes
\cite{hep-ph/9403226,hep-th/0412103,%
hep-ph/0609007,arXiv:0803.4180}. Full NLO
distributions for a growing number of $2\rightarrow4$~%
\cite{arXiv:1004.1659,arXiv:0907.1984,arXiv:0902.2760,arXiv:1105.3624,arXiv:1002.4009,%
arXiv:1012.3975,arXiv:0905.0110,arXiv:1104.2327} and even $2\rightarrow5$
\cite{arXiv:1008.5313,arXiv:1009.2338,arXiv:1108.2229} processes have now been
achieved. The degree of automation to which the virtual corrections to NLO observables can be computed
has steadily improved
\cite{arXiv:0805.2152,arXiv:0812.2998,arXiv:0911.1962,arXiv:1006.0710,%
arXiv:1103.0621,arXiv:1110.1499,arXiv:1111.2034,arXiv:1111.2708}.

In this article we review the computations of multi-parton amplitudes in
massless QCD using the \NGluon c++ library \cite{Badger:2010nx}. The algorithm employs the
generalised unitarity cutting procedure to construct one-loop amplitudes from
on-shell tree-level building blocks which we review in Section
\ref{sec:GUreview}. We then describe the inclusion of amplitudes with multiple
fermion pairs in Section \ref{sec:MultiFermions} and present
a detailed analysis of the performance and validity of our approach in Section \ref{sec:performance}.
We finish by presenting our conclusions in Section \ref{sec:conclusions}.

\section{Review of Generalised Unitarity \label{sec:GUreview}}

The method of generalised unitarity has been studied
extensively over the last few years. Based on a purely algebraic
approach, the method can be used to formulate a numerical algorithm for
the computation of one-loop amplitudes \cite{arXiv:0708.2398,arXiv:0801.2237,arXiv:0803.4180}.
Detailed and helpful reviews on the subject can be found in
references \cite{Britto:2010xq,Ellis:2011cr,Ita:2011hi}.

The one-loop amplitudes we consider in this article are colour ordered QCD
primitive amplitudes where both colour generators and internal colour flow
structure have been stripped off leaving the simplest gauge invariant 
building blocks of the full amplitude. Each of these terms has a well defined ordering of
external legs and internal propagators.

A general one-loop primitive amplitude with no external massive particles,
regulated in $D=4-2\eps$ dimensions, can be expressed in terms of a basis of
scalar two-, three- and four-point integral functions together with rational terms \cite{arXiv:0801.2237},
\begin{align}
  &A^{(1)}(p_1,\cdots,p_n) = 
  \sum_{i<j<k<l} c_{4;i,j,k,l} I_4(s_{i,k-1},s_{j,l-1};s_{i,j-1},s_{j,k-1},s_{k,l-1},s_{l,i-1})
  \nonumber\\&
  +\sum_{i<j<k} c_{3;i|k|k} I_3(s_{i,j-1},s_{j,k-1},s_{k,i-1})
  +\sum_{i<j-1} c_{2;i|j} I_2(s_{i,j-1})
  \nonumber\\&
  -\frac{1}{6}\sum_{i<j<k<l} c_{4;i,j,k,l}^{[4]}
  -\frac{1}{2}\sum_{i<j<k} c_{3;i,j,k}^{[2]}
  -\frac{1}{6}\sum_{i<j-1} s_{i,j-1} c_{2;i,j}^{[2]}
  +\mathcal{O}(\eps).
  \label{eq:1loopbasis}
\end{align}
Here $s_{i,j}=(p_i+\cdots+p_j)^2$ are the Lorentz invariants of the amplitude.
$I_4,I_3$ and $I_2$ are the well known scalar integral functions that can be
written in terms of logarithms, dilogarithms and the dimensional regulator,
$\eps$. Thanks to public programs for the
numerical evaluation of the scalar integrals for arbitrary kinematics, 
e.g. FF \cite{NIKHEF-H-90-15}, OneLOop \cite{vanHameren:2010cp} or
QCDLoop \cite{Ellis:2007qk} \footnote{ QCDLoop is the default choice in
\NGluon}, the only process dependent quantities are the coefficients
$c_{k;X}$.

The box coefficients, $c_{4;X}$, can be extracted by applying maximal cuts to
the one-loop primitive amplitude, factorises into a product of four tree-level
amplitudes. The lower point integral coefficients are then computed
systematically by considering fewer cuts and subtracting the contribution from
the higher point functions previously evaluated. In each case the integrand can be parametrised
by a polynomial of the scalar products between
the loop momentum and spurious vectors that can be neatly
described by the van Neerven-Vermaseren basis. The construction in \NGluon follows the
description of reference \cite{Ellis:2007br}. The coefficients of the 
aforementioned polynomial can be efficiently computed using a discrete Fourier projection.

For the rational terms we need to consider cuts in five, or more, dimensions. In
the Four Dimensional Helicity~(FDH) scheme this can be achieved using a mass-shifted
representation of the amplitude, where the coefficients $c_4^{[4]},c_3^{[2]}$
and $c_2^{[2]}$ can be extracted from a discrete Fourier projection over
the additional mass parameter \cite{Badger:2008cm}. The mass shifted integrands
factorise into tree-level amplitudes with massive scalar particles and massive
fermions.

\section{Extension to Multiple Fermion \label{sec:MultiFermions}} 

The extension of the pure gluonic case to multiple fermions has basically two
fundamental new features: 
\begin{enumerate} 
  \item more complicated tree-level amplitudes 
  \item restrictions when sewing tree-level amplitudes 
        together to reconstruct the integrand 
\end{enumerate}

In the pure gluonic case, all tree-level amplitudes that can occur in a
unitarity cut can be assigned a unique colour ordering. This means that they
can always be computed with the well known Berends-Giele recursion relations
\cite{Berends:1987me}. The simple reason for this is the fact that restoring the
colour to the primitive amplitude both external legs and loop lines live in the
adjoint representation. 

\begin{figure}
  \includegraphics*[width=0.33\textwidth]{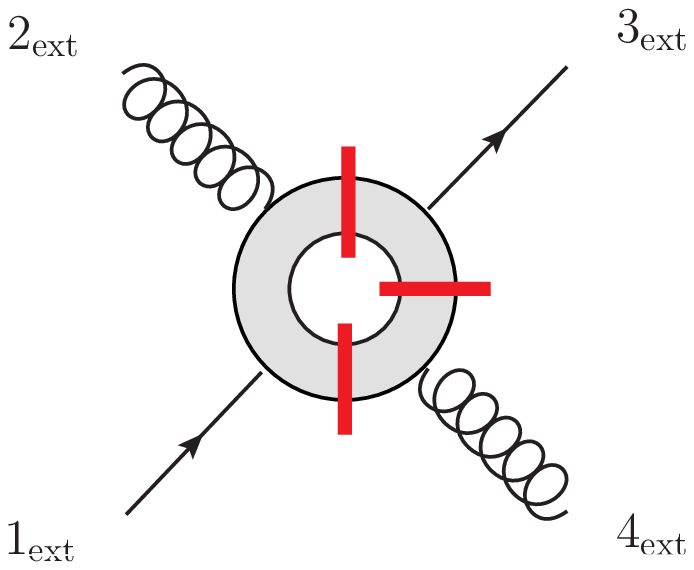}
  \includegraphics*[width=0.33\textwidth]{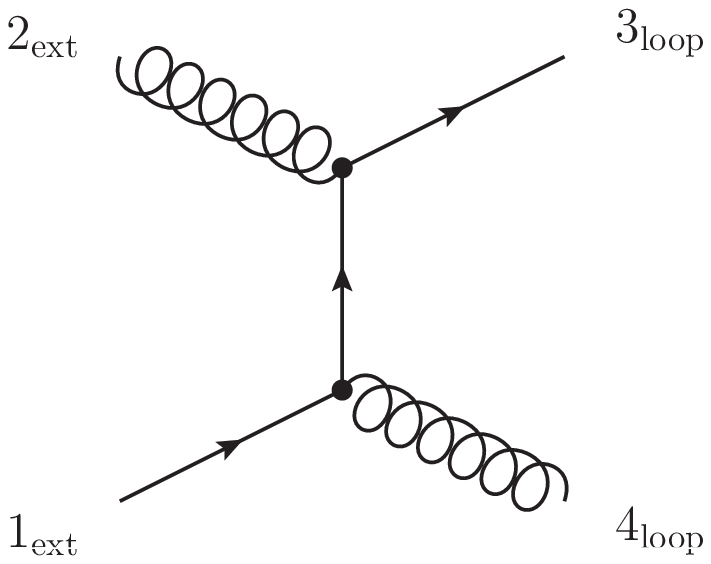}
  \includegraphics*[width=0.33\textwidth]{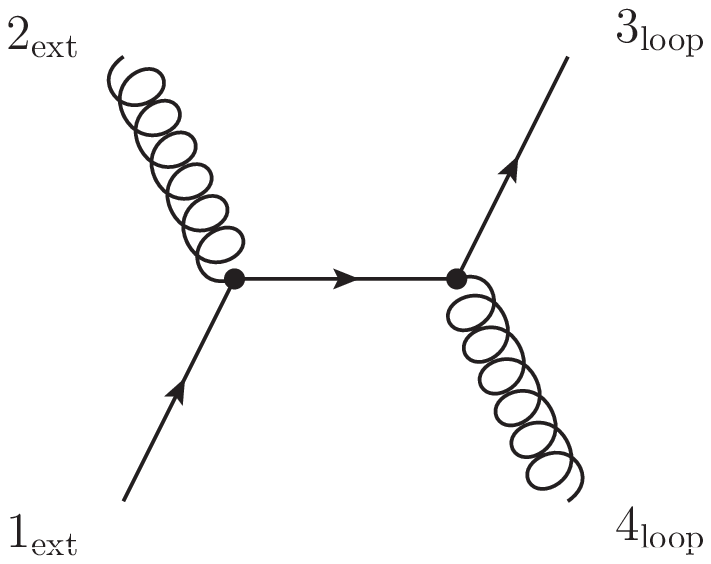} \caption{A
  triple cut of the primitive amplitude $A^{\rm prim}(\overline
  Q_1;g_2;Q_3;g_4)$ leading to a four point tree amplitude $A^{\rm
  tree}(\overline Q_{1\,{\rm ext}};g_{\,2\,{\rm ext}};Q_{\,3\,{\rm
  loop}};g_{4\,{\rm loop}})$ consisting of the above two ordered tree diagrams.
  ``ext" resp. ``loop" denote external legs, resp. on-shell loop legs. Note
  the absence of any gluon self interactions.} \label{qbgqg-example}
\end{figure}

With the inclusion of quarks in the fundamental representation, this one-to-one
correspondence is in general lost. It happens that through unitarity cuts basic building blocks appear that do not belong to any colour structure which would occur in the
computation of a Born amplitude. This can be seen, for instance, at those
quark-gluon primitive amplitudes contributing solely to the subleading colour
part of the full amplitude. As an example, consider the primitive amplitude
$A^{\rm prim}(\overline Q_1;g_2;Q_3;g_4)$ where the fixed ordering of external
particles is such that a quark anti-quark pair is separated by one gluon.  In
certain bubble and triangle cuts, for example, among others, the
following building block arises: $A^{\rm tree}( \overline Q_{1}; g_{\,2}; Q_{\,\rm
loop};g_{\,\rm loop})$ as illustrated in figure \ref{qbgqg-example}. In $A^{\rm tree}( \overline Q_{1}; g_{\,2}; Q_{\,\rm loop};g_{\,\rm loop})$, the non-abelian gluon interactions are missing and in addition,
the structure is not colour-ordered. 
It is possible to compute these building blocks as a linear combination of colour-ordered tree amplitudes. However, in
\NGluon all tree amplitudes are computed directly with help of
colour stripped Feynman rules \cite{Bern:1994fz}. One has to specify only the external leg order and
its particle content. Tree amplitudes are then evaluated independently on
whether they can be assigned a definite colour structure or not.

Technically, we follow for the trees an off-shell bottom-up approach. This
means that after the order of $n$ external legs with appropriate particle
content has been fixed, one starts computing all one-point currents, i.e.
polarisation vectors for gluons and spinors for quarks. Subsequently, higher
point currents are computed systematically from lower point currents. The
$n$-point amplitude is computed by contracting the $(n-1)$-point
current with the $n$th one-point current. This procedure has
the advantage that many currents can be reused
during the computation. The most complicated
vertex, the four gluon vertex, dictates the asymptotic scaling behaviour, which
can be shown to be polynomial of order $O(n^4)$.

In the context of generalised unitarity, an additional reduction in complexity
can be achieved: For different topologies, those currents that involve
exclusively external legs never change and can be computed once in the
beginning for all possible cuts. An amplitude evaluation is then equivalent to
joining new currents that involve loop legs with external, already known
currents. This procedure reduces the polynomial scaling even further to order
$O(n^3)$ \cite{Badger:2010nx}.

In the highly symmetric pure gluonic case, every cut is possible. In other
words, there is always a loop propagator which connects directly two external
legs. This is not necessarily the case if more then one fermion line is
involved in the amplitude. Take for example a closed quark loop and one
external $\overline q q$-pair: The external quark line will never enter the loop
and can't be put on shell.

In order to describe this problem algorithmically, we use the concept of a
parent diagram. In our framework, a parent diagram of an $n$-point amplitude is
an abstract one-loop diagram with $n$ fictitious three-point vertices and $n$
propagators. The particle content of its propagators labels the possible
on-shell settings. In addition, it knows also which propagators do not exist.
The propagator content of the parent diagram is therefore the backbone to
compute systematically all integral coefficients. The procedure is exactly like
in the pure gluonic case with the exception that if a topology involves a
non-existing propagator, the topology is simply skipped. 

In order to determine the parent diagram, one draws an abstract loop without
specifying any particle content yet and attaches all ordered external legs to
it. One starts with the specification of the particle
content of an arbitrary initial propagator in the loop and determines with the
external leg attached to this propagator the content of the next propagator in
the loop. This procedure is repeated for the whole diagram until one hits the
initial propagator.  

\begin{figure} 
 \centering
 \includegraphics[width=0.5\textwidth]{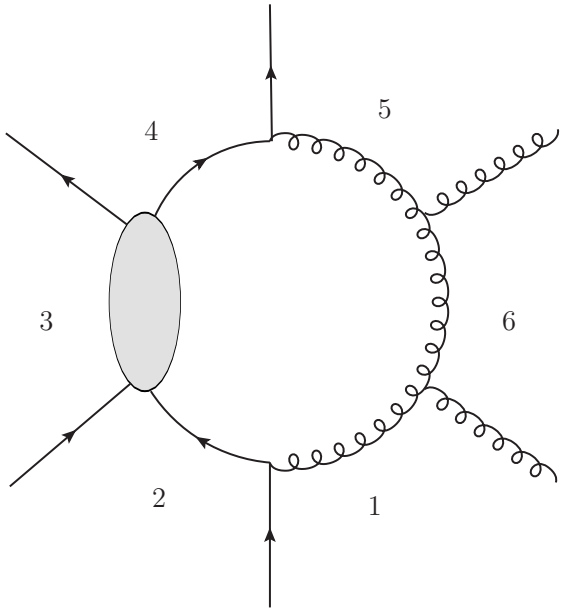} 
 \caption{A six-point parent diagram for a primitive with two external fermion lines. The
 first propagator is initialised with a gluon. The third one is then
 unphysical and represented by a blob.} \label{parentDiagram-example}
\end{figure} 

One sees immediately that certain propagators of the parent diagram can't be
assigned properly in case we are dealing with either 1) a closed fermion loop
plus at least one external fermion line or 2) two (or more) fermion lines where
a quark-antiquark pair is separated by another quark-antiquark pair. An example is given in figure \ref{parentDiagram-example}. Literally speaking, this means the following: under the grey blob in figure \ref{parentDiagram-example} live two
fermion lines which can't be represented by a valid QCD propagator. As soon as
the enclosed quark line ``leaves'' the blob, the parent diagram follows the same
pattern as before. With the knowledge of the parent diagram and the external legs, the primitive is uniquely determined.

\begin{figure} 
  \centering
  \includegraphics[width=\textwidth]{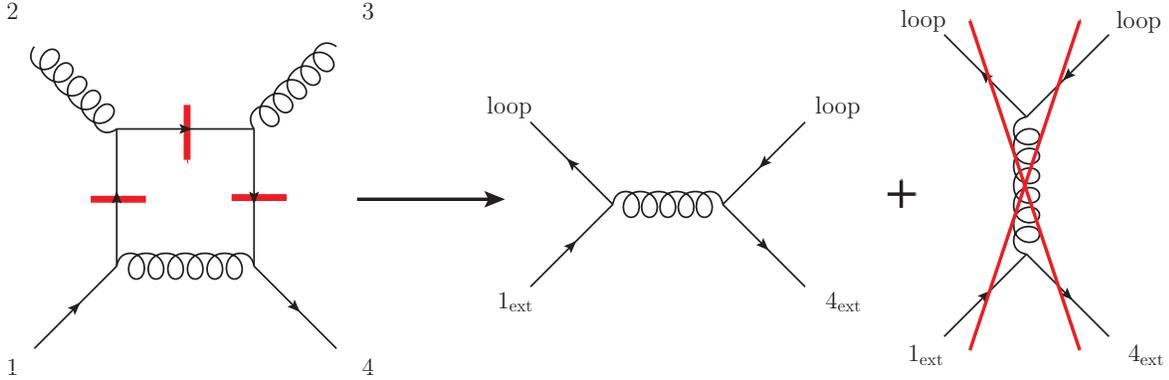}
  \caption{Triple cut with a two fermion tree amplitude. Although at one-loop
  level, there is only one fermion line, they must be treated at tree-level as
  two distinct flavours in order to avoid contributions from a closed quark
  loop (right diagram crossed out).} \label{subleadingColourAlg} 
\end{figure}

Another subtlety occurs for certain subleading colour contributions where a
distinct cut requires a tree consisting of two fermion lines which belong,
however, inside the one-loop amplitude to the same fermion line. Formally, such
trees must be treated like two different flavour lines since otherwise one
would include contributions that belong to amplitudes with a closed fermion
loop, a different class of amplitudes. A simple four-point example is given in
figure \ref{subleadingColourAlg}.

\section{Performance and Cross Checks \label{sec:performance}} 

The first detailed verification of the implementation comes from comparison
with the well known universal Infra-Red and Ultra-Violet poles in the
dimensional regulator, $\epsilon$. This analytic cross check tests the
four-dimensional (or cut-constructible) parts of the amplitude and can also
give some hints as to a loss of numerical precision.

In order to get an overall estimate for the numerical accuracy, we use the
following observation: A re-scaling of the external momenta is equivalent to a
simple change of units and should therefore be physically equivalent. Since the
floating point arithmetic at the hardware level changes, the difference between
the two evaluations --- rescaled and un-rescaled --- can be used to estimate
the numerical uncertainty. The validity of this approach is investigated in
detail in \cite{Badger:2010nx}. This test is very convenient to detect
numerical instabilities and has in addition the advantage that it is
independent from any analytic input. Of course, it is not at all suited to
check the correctness of the implementation itself and has the drawback of a factor of two in the runtime. 
In case, the internal accuracy checks do fail, the phase space point can be reprocessed using higher precision with
the qd package described in reference \cite{Hida:2008}.

\begin{figure}
  \includegraphics[width=0.5\textwidth]{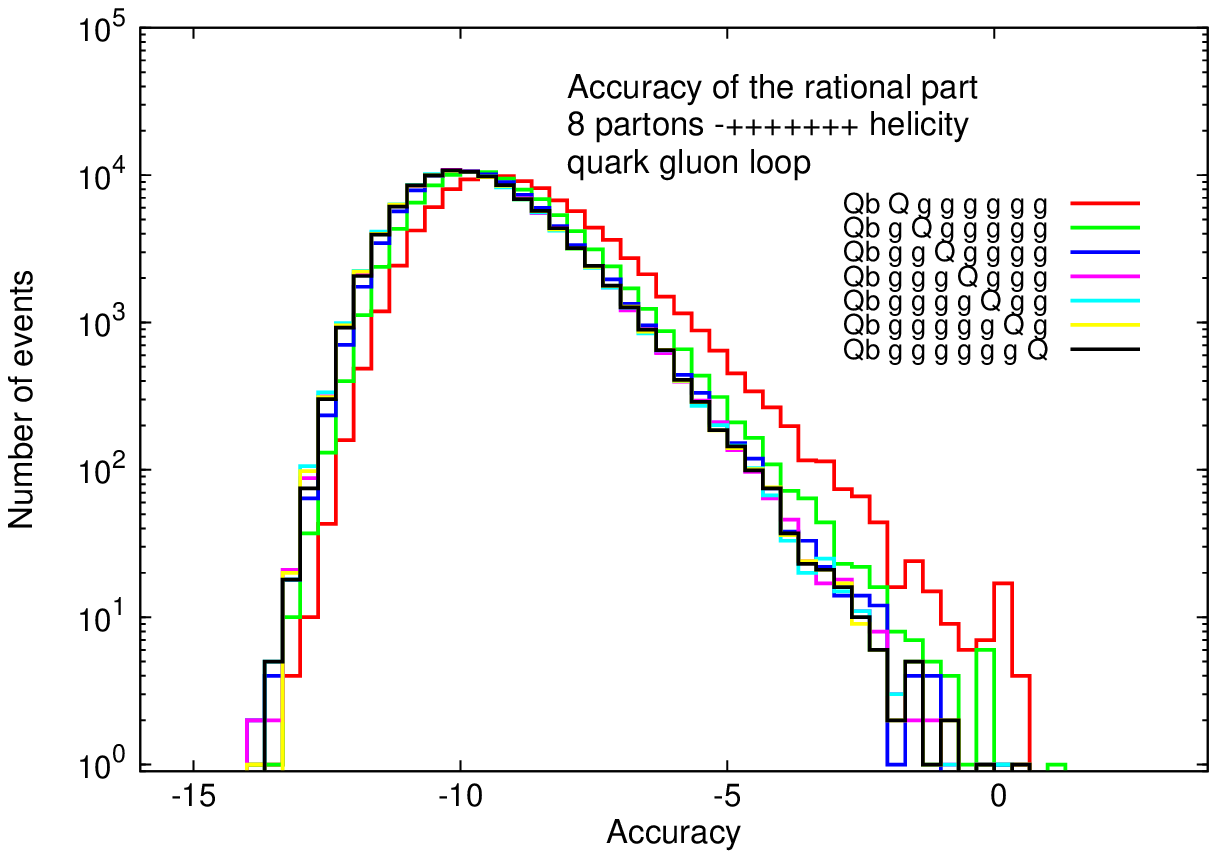}
  \includegraphics[width=0.5\textwidth]{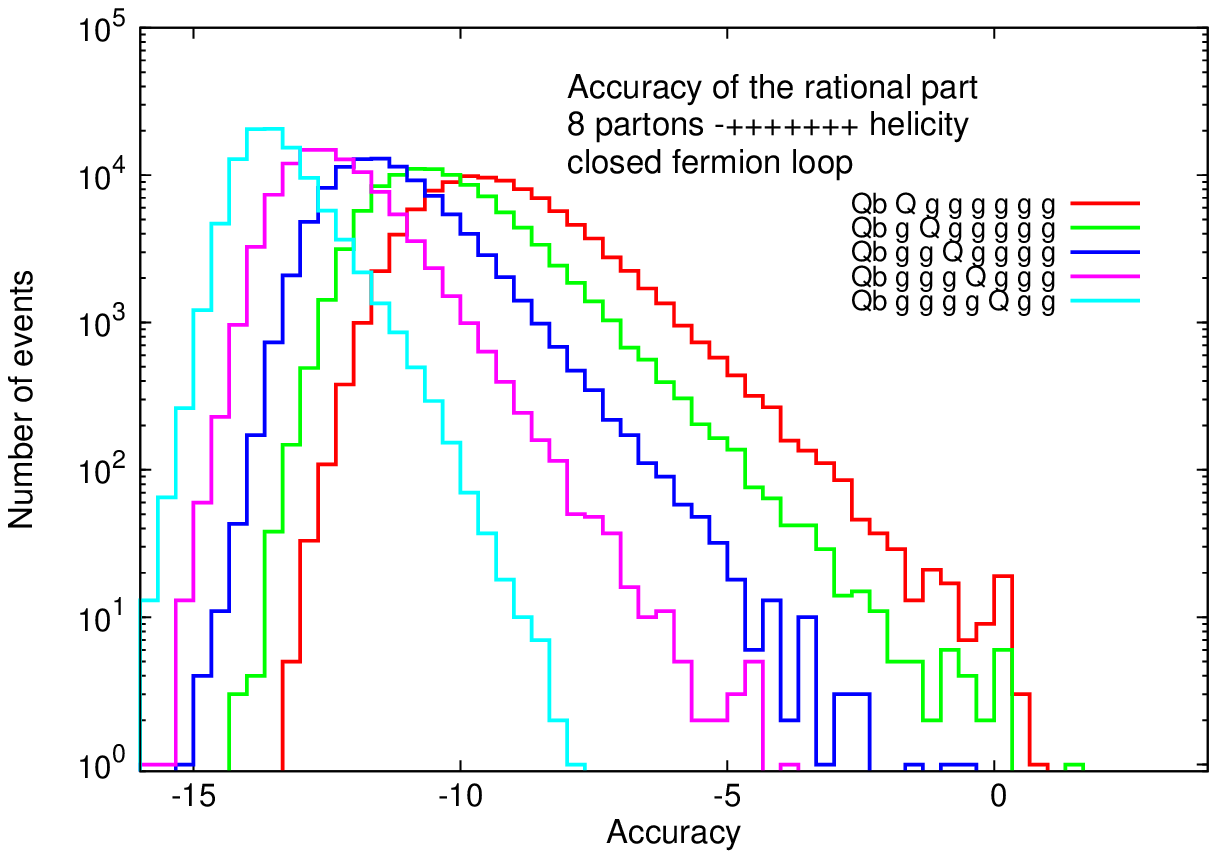}
  \caption{Accuracy distribution for the rational part of $(\overline q
  q+6\,g)$-primitives. The histograms plot the number of events against the
  decadic logarithm of the relative accuracy (\# of valid digits) between the
  result from \NGluon and the analytic formulae in \cite{Bern:2005ji} for all
  possible separations of the two external quark legs.} \label{qbqNgluon}
\end{figure}

Besides the poles, we made cross checks against known analytical and numerical
results. For the trivial extension of the pure gluonic case
to a closed quark loop with exclusively external gluons, we find full agreement
with the reference phase space points in \cite{vanHameren:2009vq}. The
correctness of the implementation for primitives with one external quark line
has been tested intensively against all formulae of the $\overline q qggg$
results given in \cite{Bern:1994fz}. The agreement of these 5-point cases holds
both for different helicity configurations and quark-antiquark separations. For
higher $n$-point functions, we checked the IR-finite amplitudes $A(\overline
q_1,g_2,...,q_i,...g_n)$ with helicity configuration $-++...+$
given analytically in \cite{Bern:2005ji} for all possible quark-antiquark separations and
find numerical agreement. An example for an 8-point function is shown in figure
\ref{qbqNgluon}. Only a very small fraction of events show an accuracy of less then 3 digits which is enough for most practical calculations. While the accuracy for different quark-antiquark separations
does not differ much in the mixed quark-gluon loop, it does for a closed quark
loop. This is simply a consequence of the fermion loop primitive amplitudes having
fewer allowed propagators as the $\overline qq$-separation increases.
This is also the reason why the computation time differs by roughly one order of
magnitude depending on the separation of the quark and the antiquark. An
example for the runtime difference is shown in table \ref{runtime}. Due to the
vanishing of massless tadpoles, the largest two separations in the closed
quark loop do not contribute.

\begin{table}
\caption{\label{runtime}Runtime in milli seconds for primitives with one quark antiquark pair
on an Intel(R) Core(TM)2 Duo CPU E8400 @ 3.00GHz processor. While in the mixed
quark-gluon loop the runtime increases with larger $\overline q q$-separation,
the contrary is the case for the closed quark loop.}
~\\[-7mm]
\begin{center}
\begin{tabular}{llllll}
  \emph{6-point example:}&&&&&\\
  \br
  $\overline q q$-separation & 0   & 1   & 2   & 3   & 4   \\
  \mr
  mixed quark-gluon loop & 4.85 & 5.95 & 7.27 & 8.77 & 10.30 \\
  closed quark loop & 6.13 & 2.61 & 0.72 & 0.01 & 0.01 \\
  \br
\end{tabular}
\end{center}
\begin{center}
\begin{tabular}{llllllll}
  \emph{8-point example:}&&&&&&&\\
  \br
  $\overline q q$-separation & 0   & 1   & 2   & 3   & 4 & 5 & 6  \\
  \mr
  mixed quark-gluon loop & 19.84 & 22.08 & 23.81 & 27.35 & 30.54 & 34.13 & 38.07 \\
  closed quark loop & 26.31 & 14.82 & 7.57 & 3.14 &  0.91 & 0.03 & 0.03 \\
  \br
\end{tabular}
\end{center}
\end{table}

For the multiple fermion case, we find agreement in the poles for all
primitives that we have tested (up to 5 external quark-antiquark pairs). 

When computing the full amplitude out of primitives, many
tree-level amplitudes can be reused. The necessary conditions are: 1) the loop
momentum for a topology agrees, 2) both loop and external flavours agree and 3)
both loop and external helicities agree. It is important to stress that for one
primitive with fixed helicity, no information can be reused. The cache system
starts first its work when one is dealing with either different permutations of
primitive amplitudes or with different helicity amplitudes (if, for example, an
average over all helicities is carried out). Take for instance a bubble cut
with two trees: flipping one helicity in one tree does not affect the other
tree which can be recycled. Similar considerations apply to permutations that
affect only trees inside one cut: At any time, the other tree is left
untouched and can be reused. Note that we reuse only full tree amplitudes and no off-shell
currents. Especially when dealing with permutations, a cache that works on
current level could be constructed. The book-keeping for such an approach is
however much more involved. As a consequence this possibility is not yet used
in \NGluon.

We have implemented the cache system via a binary tree which is part of the c++
standard template library. Due to the internal index system, the cache is
restricted to 13 external legs\footnote{This is restricted to 6 legs on 32 bit machines} with six (different flavour) external quark
pairs. The asymptotic behaviour is estimated via the computation time of
$(n-1)!/2$ permutations of primitives. For low multiplicity (in our case four
point amplitudes), there is an overhead using such a system since the stored
amplitudes are usually only three-point or at most four point functions and
very quickly evaluated. From five external legs on until eight one gains speed
up factors between 2 and 3. For higher multiplicities, however, the amount of
storage information is so large that the cache starts to slow down the
computation compared to the direct evaluation without cache. It is expected
that for the leading colour approximation where far less primitives are needed,
the performance of the cache is more efficient.

\section{Conclusions \label{sec:conclusions}}

\NGluon is a publicly available program to compute primitive amplitudes with a
fixed order of external legs in pure gauge theory. We have shown the extension
of the program to an arbitrary number of external massless quarks: For a fixed
order of helicity and flavour of the external particles, the primitive
amplitude is evaluated in a fully automatic way. As internal accuracy check, we
find that the ``scaling test'' gives reliable estimates. We have made several
cross checks with known results from the literature and find numerical
agreement with our implementation. An additional cache system which recycles tree
amplitudes for the unitarity cut can lead to speed up factors
between two and three. First phenomenological applications are in preparation.

\end{document}